\documentclass[proceedings, preprint]{rmaa}

\usepackage{paralist}
\usepackage{psfrag,color}

\SetYear{2007}
\SetConfTitle{XII Reuni\'on Regional Latinoamericana de la UAI}

\title{On the variability of HD 170699 -- a possible COROT Target}    

\author{
 M. Alvarez\altaffilmark{1,8}, JP. Sareyan\altaffilmark{2}, L. Parrao\altaffilmark{3}, 
JH. Pe\~na\altaffilmark{3}, L. Fox-Machado\altaffilmark{1}, E. Poretti\altaffilmark{4}, 
S. Martin-Ruiz\altaffilmark{5}, P. Amado\altaffilmark{5}, \  R. Garrido\altaffilmark{5}, 
C. Aerts\altaffilmark{6}, Z. Csurby,\altaffilmark{7} and M. Paparo,\altaffilmark{7}}

\altaffiltext{1}{Observatorio Astron\'omico Nacional, Instituto de Astronom\'{\i}a, 
Universidad Nacional Aut\'onoma de M\'exico, Apto.Postal. 877, Ensenada, 
BC 22860,  M\'exico   (alvarez@ astrosen.unam.mx).}
\altaffiltext{2}{Observatoire de la Cote dAzur, GEMINI, UMR 6203, BP 4229, 06304
Nice Cedex 4, France.}
\altaffiltext{3}{Instituto de Astronom\'{\i}a-Universidad Nacional Aut\'onoma de 
M\'exico,  M\'exico.}
\altaffiltext{4}{INAF Osservatorio Astronomico di Brera,Via Bianchi 46, 23807 Merate,
Italy.}
\altaffiltext{5}{ Instituto de Astrof\'{\i}sica de Andaluc\'{\i}a, Consejo Superior 
de Investigaciones Cient\'{\i}ficas, Apartado 3004, 18080 Granada, Spain.}
\altaffiltext{6}{Instituut voor Sterrenkunde, KatholiekeUniversiteit Leuven, 
Celestijnenlaan 200 B, 3001 Leuven, Belgium.}
\altaffiltext{7}{Observatory, Hungary.}

\shortauthor{Alvarez, Sareyan, Parrao, et al.}
\shorttitle{Variability of HD 170699 -- a COROT Target}

\listofauthors{M. \'Alvarez, \ JP. Sareyan, \ L. Parrao, \ JH. Pe\~na, \ L. Fox,
 \ S. Martin, \ P. Amado, \ R. Garrido, \ C. Aerts, \ Z. Csurby, \ M. Paparo.}

\indexauthor{\'Alvarez, M.}
\indexauthor{Sareyan, JP.}
\indexauthor{Parrao, L.}
\indexauthor{Pe\~na, JH.}
\indexauthor{Fox-Machado, L.}
\indexauthor{Poretti, E.}
\indexauthor{Martin, S.}
\indexauthor{Amado, P.}
\indexauthor{Garrido, R.}
\indexauthor{Aerts, C. }
\indexauthor{Csurby, Z.} 
\indexauthor{Paparo, M.}

 \abstract{We present the analysis of the variability of HD 170699, a COROT star 
showing the characteristics of a non evolutionary $\delta$ Scuti star with high rotational 
velocity.   There is a clear period of 10.45 c/d with 5.29 mmag amplitude in the $y$
filter.   From the data, it can be seen that the star shows multi-periodicity and it is 
necessary to add more frequencies to adjust the observationas.} 

 \resumen{Se presenta el an\'alisis de la variabilidad de HD 170699, una estrella del
 programa COROT que muestra ca\-racter\1sticas de una estrella tipo $\delta$ Scuti
 no evolucionada con alta velocidad de rotaci\'on. \ Muestra una  variabilidad con un 
 claro per\1odo de 10.45 c/d con 5.29 mmag de amplitud en el filtro $y$. \ De las 
 observaciones, podemos ver que la estrella es multi-peri\'odica y que se requiere 
 a\~nadir mas frecuencias para ajustar las observaciones.}

\addkeyword{Stars: COROT Target}
\addkeyword{Stars: $delta$ Scuti: HD 170699}
\addkeyword{Stars: HD 170699}
\addkeyword{Stars: Variable stars}

\begin{document}
\maketitle

\section{General}
\label{sec:intro}

The study of the interior of the stars is a well established branch of $asteroseismology$.  Photometric studies of $\delta$ Scuti stars, $\gamma$ Dor, $\beta$ Ceph among 
others has shown to be a very good tool to determine the oscilatory behaviour 
of  these interesting objects.  Recently, within the framework of COROT`s program, a 
group of stars were observed from different ground based observatories to select 
some of the {\it primary} and {\it secondary} targets, both in the $Galactic$ and 
$Antigalactic \ Center$ direction within the field of view of the satellite.  Preliminary 
results of COROT`s  program, are in the e-page  http://exoplanet.eu/corot.html

HD 170699 is an A2 star in a double system with magnitude B=7.44 and V=6.956;  it 
was reported as a possible multiperiodic variable by Poretti et al. (2003) after one 
night of observations in 2002. \ The position of HD 170699 within the field of view 
of COROT`s satellite, makes it a very promising object to study.  Its high rotational 
velocity of v sin$i$ $ >$ 200 km s$^{-1}$, makes it an interesting object to  
understand its behavior within the pulsation treatment. 

\begin{figure}[!t]
  \includegraphics[width=\columnwidth]{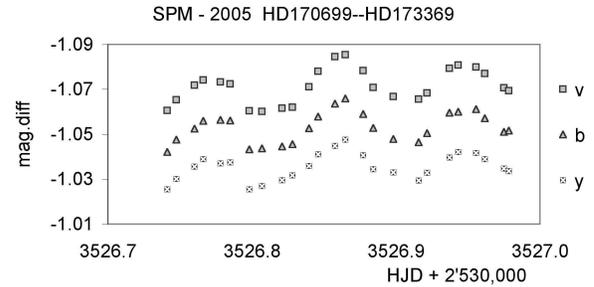}
  \caption{Differential photometry of HD170699--HD173369 during the HJD 2453,526 
at  SPM observatory. A period between 1.6 to 2.1 hours can be observed from our
measurements.}
  \label{fig:alvarezmp1}
\end{figure}

 \section{{Observations -- photometry.}}
 \subsection{Differential photometry.}
\label{sec:diffot}

As a continuation of monitoring some $\delta$ Scuti and $\gamma$ Dor objects,  
HD 170699 (HIP 90728, SAO 123585), a $secondary \ target$ of COROT`s program
was observed  at San Pedro M\'artir Mexican observatory (SPM) and Sierra Nevada  
observatory (OSN), Granada, Spain with the $uvby$ Str\"omgren system during 
34 nights on 2007, 2005, 2004 and 2002, making more than 180 hours of observation.  
During the 2007 season, we included the H$_{\beta}$ filter in our measurements. 
The star was also observed at Hungary and by the MERCATOR team.

Figure~\ref{fig:alvarezmp1} shows the differential photometry of 
HD 170699--HD 173369, during one night at SPM Mexican observatory.  
A clear variation with three maxima and possible multiperiodicity is seen in 
filters $(v, b, y)$, during the almost 6 hours of observation.

As comparison stars, we observed HD 173369 and HD 172046 and finally we 
chose the first  HD 173369, for all the program, except for the first night 
in 2002, were HD 166991 was the only comparison star of the observing run.

\begin{figure}[!t]
  \includegraphics[width=\columnwidth]{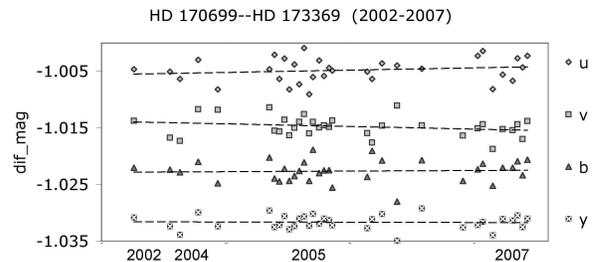}
  \caption{Average daily values of the differential photometry of HD 170699-HD173369
from 2002 to 2007. } 
  \label{fig:alvarezmp2}
\end{figure}

In Figure~\ref{fig:alvarezmp2} we show the average daily values of Str\"omgren 
filters $u, v, b, y$ for the length of the campaign, we arbitrary displaced in time to 
show the different epochs of observation.    The daily change give us a clear 
indication of the variability in this star.   Filter $u$ shows a strong instability in their 
values and we did not include this filter in our analysis. 


\subsection{Absolute photometry.}
\label{sec:absfot}

We report our system photometry observations obtained during the 2007 season, 
we observed the variable and the comparison stars with the Str\"omgren system 
including the H$_{\beta}$  filter.  During the campaign, we observed more than 39 
standard stars to obtain the absolute calibration of them.  \ 

In Table~\ref{tab:own_measure} we report our own measurements and  those given at 
the CDS (Simbad) for HD 170699 and the comparison stars used during 2004 to 2007 
campaign at SPM and OSN observatories.

Mercator team measured on the {\it Geneva} system and their results will be given 
elsewhere.  The two nights measurements of M. Paparo and  Z. Csurby were done in 
the V filter and we are not using their results in this work. \ The averaged instrumental 
value they obtain from two nights of measurements is  V = $-1.026$   

\begin{table}[!t]\centering
  \setlength{\tabnotewidth}{0.4\textwidth}
  \setlength{\tabcolsep}{0.4\tabcolsep}
  \tablecols{7}
  \caption{HD 170699 and program stars} 
  \label{tab:own_measure}
  \begin{tabular}{l c c c c c  r}
  \toprule
 Star  &  V    &  b-y     &  m$_{1}$ & c$_{1}$ & H$_{\beta}$   & \# mes \\ 
    \midrule
170699  & 6.959  & 0.127  & 0.163  & 0.978 & 2.796   & 625/36  {$\dagger$} \\ 
173369 &  7.989  & 0.115  & 0.232  & 0.887  & 2.839  & 650/42  {$\dagger$} \\ 
172046 &  6.671 & 0.035  & 0.071  & 0.551  & 2.695   & 642/38  {$\dagger$} \\ 
    \midrule
170699 &  6.956  & 0.127  & 0.174  & 0.985  & 2.820  &  2/2  {$^{*}$} \\ 
173369 &  7.970  & --        & --        & --        & --          & --       {$^{*}$} \\ 
172046 &  6.676  & 0.038  & 0.065 & 0.611  & 2.704   & 12/10  {$^{*}$} \\ 
  \midrule
 \tabnotetext {$\dagger$}{This paper -- 39 standard stars observed during our 
 June-July 2007 measurements.}
 \tabnotetext {$^{*}$}{ CDS Simbad}
\end{tabular}
\end{table}

\section{{Frequency analysis.}}

For the frequency analysis, we used PERIOD04 (Lenz and Breger, 2005) numerical 
package and PERANSO software (Vanmunster, T., 2007). We analyzed the whole set of
data, we used SPM data (9 nights in 2007, 13 nights in 2005 and 2 nights in 2004) 
and  also include OSN data (7 nights in 2005, 2 nights in 2004).

In Table~\ref{tab:freq1} we report three frequencies found for this star. There is a 
clear frequency of 10.45 c/d with amplitude of 5 to 6 mmag.  Other frequencies at 
2.46, 3.48, 4.45 and 3.18 c/d with amplitudes of 2 and 3 mmag may be present, 
although there is not enough information to discriminate between them.  Clearly, it is 
a good target for COROT`s program.  The final analysis of this work is under 
preparation and will be published elsewhere.

{\bf This research was done as part of PAPIIT IN108106 Program of UNAM.}

 \begin{table}[!t]\centering
 \setlength{\tabnotewidth}{0.45\textwidth}     
 \setlength{\tabcolsep}{0.5\tabcolsep}
 \tablecols{5}
 \caption{Frequency of HD 170699  --  2002 to 2007} 
 \label{tab:freq1}
 \begin{tabular}{l l l l l }
    \toprule
filter & f${_1}$ ( A${_1}$) & f${_2}$ ( A${_2}$) &  f${_3}$ ( A${_3}$) \\
 \midrule 
  $v$ & 10.4509(6.88) & 2.4604(3.37) & 3.1904(3.14) \\
  $b$ & 10.4509(6.54) & 3.4844(3.33) & 3.1809(3.01) \\
  $y$ & 10.4509(5.29) & 3.1838(2.26) & 4.4539(1.97) \\
 \midrule 
\tabnotetext {$*$}{$u$ measurements show a poor signal  (not used in the analysis) --
   freq. in c/d -- amp in mmag.}
\end{tabular}
\end{table}

\end{document}